\documentstyle[12pt]{article}
\begin{document}
\baselineskip=0.7cm
%
\def\half       {  {1\over 2}  }
\def\ie         {  {{\it i.e.}\  }    }
\def\Tr         { {\rm Tr}\,  }
\def\com#1#2   { \left[#1, #2\right]}
\def\comma          {\, ,}
\def\period         {\, .}
\def\nn               {  \nonumber  }
\def\calO{{\cal O}}
\def\calL{{\cal L}}
\def\del{\partial}
\def\al{\alpha}
\def\be{\beta}
\def\ga{\gamma}
\def\ep{\epsilon}
\def\lam{\lambda}
\def\Ga{\Gamma}
\def\de{\delta_\ep}
\def\dme{\delta^{(M)}_\ep}
\def\De{\Delta_\ep}
\def\dbe{\bar{\delta}_\ep}
\def\dte{\tilde{\delta}_\ep}
\def\dk{\delta_K}
\def\cbar{\bar{C}}
\def\Atil{\tilde{A}}
\def\vtil{\tilde{v}}
\def\eptil{\tilde{\ep}}
\def\xtil{\tilde{x}}
\def\ytil{\tilde{y}}
\def\Stil{\tilde{S}}
\def\calI{{\cal I}}
\def\Xtil{\tilde{X}}

\newcommand{\EQ}{\begin{equation}}
\newcommand{\EN}{\end{equation}}
\newcommand{\EQA}{\begin{eqnarray}}
\newcommand{\EQN}{\end{eqnarray}}
\newcommand{\e}{{\rm e}}
\newcommand{\Sp}{{\rm Sp}}
\makeatletter
\def\section{\@startsection{section}{1}{\z@}{-3.5ex plus -1ex minus
 -.2ex}{2.3ex plus .2ex}{\large}}
\def\subsection{\@startsection{subsection}{2}{\z@}{-3.25ex plus -1ex minus
 -.2ex}{1.5ex plus .2ex}{\normalsize\it}}
\def\appendix{
\par
\setcounter{section}{0}
\setcounter{subsection}{0}
\def\thesection{\Alph{section}}}
\makeatother
\def\thefootnote{\fnsymbol{footnote}}

\begin{flushright}
BROWN-HET-1146 \\
UT-KOMABA/98-24\\
October 1998
\end{flushright}

\begin{center}
\Large
Generalized Conformal Symmetry \\in
D-Brane Matrix Models

\vspace{1cm}
\normalsize
{\sc Antal Jevicki}
\footnote{
{\tt antal@het.brown.edu}
}

\vspace{0.3cm}
{\it Department of Physics, Brown University\\
Providence, RI 02912
}

\vspace{0.5cm}
{\sc Yoichi Kazama}
\footnote{
{\tt kazama@hep1.c.u-tokyo.ac.jp}
}
 \quad and \quad {\sc Tamiaki Yoneya}
\footnote{
{\tt tam@hep1.c.u-tokyo.ac.jp}
}

\vspace{0.3cm}
{\it Institute of Physics, University of Tokyo\\
Komaba, Meguro-ku, 153 Tokyo}

\vspace{1cm}
Abstract
\end{center}

 We study in detail the extension of the generalized conformal symmetry
proposed previously for D-particles to the case of supersymmetric
Yang-Mills matrix models of D$p$-branes for arbitrary $p$.
It is demonstrated that such a symmetry indeed exists both
 in the Yang-Mills theory  and in the corresponding supergravity
backgrounds produced by D$p$-branes.
 On the Yang-Mills side, we derive
the field-dependent special conformal transformations
for the collective coordinates of D$p$-branes in the one-loop approximation,
and show that they coincide with the transformations on the supergravity side.
These transformations are powerful in restricting the forms of the effective
actions of probe D-branes in the fixed backgrounds of source D-branes.
Furthermore, our formalism
 enables us to extend the concept of (generalized)
 conformal symmetry to arbitrary configurations of D-branes, which
 can still be used to restrict the dynamics of D-branes. For such general
 configurations, however, it cannot be endowed a simple classical
 space-time interpretation at least in the static gauge
adopted in the present formulation of D-branes.

\section{Introduction}
Conformal symmetry has long been playing a prominent role both in the
 realm of local field theory and of string theory.
 Especially, in the latter,  the world-sheet
(super) conformal symmetry stands out as the single most important
 principle, at least in its perturbative formulation, which endows
 it  the characteristic features of a unified theory
 of all the interactions of nature including gravity.
\par
 In the recent developments in string dualities, notably in the context of
 conjectured AdS/CFT correspondence   \cite{maldacena}\cite{gklepolya}\cite{witten}, 
the conformal symmetry continues to play a pivotal role.
For example, in the prototypical case of D3-brane system, the
 $AdS_5 \times S^5$ space-time produced by a large number of
coincident D3-branes and the  ${\cal N}=4$ super Yang-Mills theory
describing the low energy  dynamics of them  share
 the same symmetry group  including the conformal group $SO(4,2)$.
This symmetry, together with supersymmetric non-renormalization
 theorems, was shown \cite{maldacena} to be powerful enough to fix
the form of the effective action for a probe D3-brane in such a space-time
 in the   near-horizon limit.
\par
      From the standpoint  of the  {\it space-time uncertainty principle}
 proposed by one of the present authors \cite{yon}, which qualitatively
 captures the essence of the short-distance space-time structure in string
 theories, the existence of  such a conformal symmetry is deeply connected to
 the opposite scaling laws for the \lq\lq longitudinal"
 coordinate (including time) $X_\parallel$ and the \lq\lq transverse" coordinate
$X_\perp$.
For the brane system in question, the former corresponds to the
 world volume space-time coordinate $x^\alpha$ while the latter refers to
 the target space (spatial)
coordinate $x^m$ transverse to the brane,  and
 the proposed principle expresses the duality between the small and
 the large distance scales for these two categories of coordinates.
This duality is at the heart of the $s$-$t$ duality, which in turn should
 be the basis of the AdS/CFT correspondence. It is also intimately related
 to the so-called {\it UV/IR correspondence} \cite{susswitt} thought to
 be the mechanism underlying the holographic principle   \cite{thsuss}.
\par
As the viewpoint described above is not restricted to any particular
 D-brane, it is  natural to suspect
 that  some form of conformal invariance might  exist for D$p$-brane
 systems for general value of $p$, not just for  $p=3$.
Indeed in a previous paper \cite{jy}, two of the present authors demonstrated
 that a conformal symmetry of $SO(2,1)$ type exits for D0-brane
 system both for the supergravity solution and for the super Yang-Mills
 matrix theory,  {\it provided} that the string coupling $g_s$
 (or the Yang-Mills coupling $g^2$) is also transformed like a
 background field of dimension $3$.
Although it is not a symmetry in the strict sense of the word, it {\it is} a
powerful structure with which one can derive Ward identities that
 govern the theory\footnote{For instance, this type of
symmetry structure is successfully used  in \cite{seiberg} to prove
 a non-renormalization theorem.}. In fact it was shown that the
 effective action for the probe D-particle in the near-core limit
 dictated by this structure coincides with the one obtained
 in Matrix theory calculation \cite{bbpt}
  in the discrete light-cone prescription.
 In this sense, the above symmetry structure should deserve to be called
 a {\it generalized conformal symmetry (GCS)}.
\par
In the same work \cite{jy}, another important aspect of
the (generalized) conformal
 invariance was recognized and emphasized.
  It is related to
the difference in the forms  of the special conformal
transformation (hereafter referred to as SCT)
 on the supergravity side and on the Yang-Mills side.
 This difference already exists in the case of D3-brane system.
 For the world-volume coordinates of the
 Yang-Mills theory, it is of  the canonical form, namely
 $\de x^\al = 2\ep\cdot x x^\al -\ep^\al x^2$. Contrarily,
 SCT that leaves the metric of $AdS_5\times S^5$ invariant  has a
 non-canonical form
\begin{eqnarray}
\de^{AdS} x^\al &=& 2\ep\cdot x x^\al -\ep^\al x^2 -\ep^\al {2g^2 N
 \over U^2} \comma \label{adssct}
\end{eqnarray}
with the last term depending both on the Yang-Mills coupling $g$ and on
  the transverse radial coordinate $U (=r/\al')$. Since $U$ corresponds to the
 expectation value of the diagonal part of the Higgs field on the Yang-Mills
 side, the latter dependence is  field-dependent as well as non-linear.
 This field-dependence is of utmost importance
in restricting the dynamics of D-branes,  as it  connects the
terms in the effective action  which,  from the viewpoint of
the Yang-Mills theory, are induced at different loop orders.
\par
This difference in realization is formally consistent in the usual
picture \cite{gklepolya}\cite{witten}
of the Yang-Mills theories, where  these
theories are considered to live on the boundaries of the AdS space-times,
since the transformation trivially reduces to the usual linear one
 as $U\rightarrow \infty$.
In this interpretation, the information of the Yang-Mills theory
is used only as the boundary condition for the theory in the bulk.
\par
However, if one takes the super Yang-Mills theory as
 the dynamical theory of D-branes, then the situation becomes
 quite different. Since  one can place the D-branes
anywhere in the bulk, in order for the D-branes
to correctly detect the supergravity effect, the non-linear
field-dependent SCT which characterizes the gravity in the bulk
 must emerge within the Yang-Mills theory. It is not at all
 evident how such a field-dependent transformation,
whose origin is the isometric diffeomorphism of supergravity,
is derived from the linear conformal transformation  of the Yang-Mills
theory. As the (generalized) conformal symmetry is the basic underlying
 structure that supports the conjectured duality, understanding
 of this problem is clearly of prime importance. In previous
 studies, including the Maldacena's original work, this issue
 however remained untouched.
\par
Very recently, we have succeeded in resolving this issue in the
 case of  D3-brane system \cite{jky}.  We showed that
SCT law for the diagonal Higgs
 field in ${\cal N}=4$ Yang-Mills theory is modified by what we called
  the \lq\lq quantum metamorphosis" effect associated with
 the loops of off-diagonal massive fields and that, to the leading
 order in the velocity expansion, it takes exactly the form of
 the AdS transformation law (\ref{adssct}), including
 the numerical coefficient. The key observation
 was that SCT changes the gauge orbit specified by a background
 gauge and the extra transformation necessary to get back to the
 original gauge orbit induces the desired correction in the transformation
 of the diagonal Higgs field.
\par
The purpose of this article is to extend the previous discussions
 to the system of D$p$-branes for general $p$. Specifically,
 we will provide the
 answers to the following questions:
\begin{description}
\item[(i)] Does there exist a generalized
conformal symmetry on both the supergravity and the super Yang-Mills side for D$p$-brane system for general $p$ ?
 \item[(ii)] If it does, how are its realizations on respective side
 related ?
\end{description}
While the affirmative answer to (i) can be obtained rather straightforwardly
 along the lines of the previous work on the  D0-brane system  \cite{jy},
 the proper understanding of (ii) turned out to involve an intriguing
 subtlety compared with the D3-brane case treated in   \cite{jky}.
\par
The organization of the rest of the article will be as follows:
In section II, we present the generalized
conformal transformations  both from  the viewpoints
of supergravity and of super Yang-Mills matrix models.
We demonstrate how GCS can be used to
determine the effective DBI (Dirac-Born-Infeld)
 actions for the probe D$p$-branes.
Section III deals with the problem (ii) stated above. We will first
 generalize the mechanism of \lq\lq quantum metamorphosis" previously
 found for D3-brane system and establish the precise form
 of {\it quantum GCS} for D$p$-brane super Yang-Mills theory. We then
 compute the form of the modified SCT for the diagonal
 Higgs field in one-loop approximation. This turned out to differ by
 a $p$-dependent factor from the one expected from supergravity.
To understand this apparent discrepancy, an explicit calculation
 of the effective action for D$p$-branes will be performed, with
 careful treatment of the dependence of the string coupling
 on the world-volume coordinates. We will find that the result contains
 an additional term  proportional to the derivative of the coupling, which
 nevertheless is completely consistent with the quantum GCS. We then
 go on to demonstrate that appropriate redefinitions of the
 collective coordinates of the D$p$-branes in the Yang-Mills theory
 remove this extra term and correct the factor in the SCT law to
 the desired value {\it simultaneously}. This mechanism will be shown
 to be understood from the supergravity side as well.
Our discussion on this point will disclose some remarkable
consistency between supergravity and super Yang-Mills
matrix models, which to our knowledge has never been
envisaged in the previous literature.
As the final topic in section III, we will briefly discuss a
generalization of our result to
 more general configurations of D-branes, taking the case of D-particles
as the simplest example.  In the concluding section,
we  discuss the remaining problems as well as
possible further implications and extensions of the (generalized) conformal symmetries.
\section{Generalized Conformal Symmetry (GCS) for D$p$-Branes}
\subsection{GCS for the metric and DBI action}
Consider the supergravity solution produced by  $N$ coincident D$p$-branes
 at the origin. The near-horizon (or more appropriately,
`near-core' for general $p$) limit
 of interest is defined by\footnote{We
follow the convention of \cite{Itzhakietal}, including the use of
 space-favored  metric $\eta_{\mu\nu} = {\rm diag}\,(-,+,+,\cdots,+)$.}
\begin{eqnarray}
\al' &\rightarrow & 0 \comma \\
g^2 &=& (2\pi)^{p-2} g_s {\al'}^{(p-3)/2} = \mbox{fixed}\comma  \\
U &=& {r \over \al'} = \mbox{fixed} \comma
\end{eqnarray}
where $g$, $g_s$, $r$ are, respectively, the Yang-Mills coupling, the string
 coupling, and the transverse distance from the branes. In this limit, the
 metric, the dilaton and the $(p+1)$-form RR gauge  fields can be written
in the following  form:
\begin{eqnarray}
ds^2 &=& \al'\left( h_p^{-1/2}dx^2
 + h_p^{1/2} ( dU^2 + U^2 d\Omega^2_{8-p})\right) \comma \\
e^\phi &=& g_s \left({h_p\over {\al'}^2}\right)^{(3-p)/4} \comma \\
A_{0\ldots p} &=& -{1\over 2g_s}
 \left({h_p\over {\al'}^2}\right)^{-1} \comma  \\
h_p &=& {Q_p \over U^{7-p}} \comma \\
 Q_p &=& g^2N d_p \comma \qquad d_p = 2^{7-2p}\pi^{(9-3p)/2}\Ga
\left( {7-p \over 2}\right) \period
\end{eqnarray}
Let us  introduce a convenient dimensionless variable $\rho_p$ defined by
\begin{eqnarray}
\rho_p &\equiv & {Q_p \over U^{3-p}} \period
\end{eqnarray}
Then the  metric  can be written in a suggestive form as
\begin{eqnarray}
ds^2 &=& \al'\left( {U^2 \over \sqrt{\rho_p}}dx^2
 + {\sqrt{\rho_p}\over U^2}dU^2
 + \sqrt{\rho_p} d\Omega_{8-p}^2 \right) \period \label{dpmetric}
\end{eqnarray}
Except for $p=3$, $\rho_p$ is coordinate-dependent and hence the
 space-time is not exactly of AdS type. But if $\rho_p$ were constant,
 the metric would be that of $AdS_{p+2}\times S^{8-p}$ and this prompts us
 to seek a generalized conformal transformation that {\it leaves
 $\rho_p$ invariant.}
\par
Since the scale and the Lorentz invariance are trivial, we will concentrate
 on the special conformal transformation. Take the usual transformation
 law for the variable $U$, namely,
\begin{eqnarray}
\de U = -2\ep\cdot x\, U \period
\end{eqnarray}
Then, the requirement $\de \rho_p=0$ readily leads to
\begin{eqnarray}
\de Q_p &=& -2(3-p)\ep\cdot x \, Q_p \period
\end{eqnarray}
This means that we must treat  $Q_p$ (\ie $g_s$) not as a
 strict constant but  as a \lq\lq field"
on the world-volume, to the linear order in $x$, before making
  SCT. Once SCT is made, we may set it to a constant.
As for the transformation of $x^\al$, we assume the AdS-like form
\begin{eqnarray}
\de x^\al &=& 2\ep\cdot x x^\al-\ep^\al x^2
 -\ep^\al{k\rho_p\over U^2}\comma \label{dexads}
\end{eqnarray}
with some constant $k$.  It is then straightforward to check that the
 metric (\ref{dpmetric}) is invariant under the SCT defined above,  provided
we take
\begin{eqnarray}
k &=& {2\over 5-p} \period
\end{eqnarray}
This indeed covers the D0-brane case previously
studied  in \cite{jy} as well as the D3-brane case.
\par
Let us now  demonstrate that  GCS  governs the DBI effective action
 for a radially moving probe D$p$-brane in the field of a heavy source
 consisting of $N$ coincident D$p$-branes placed at the origin.
Rather than checking the invariance of the DBI action directly, it is
 instructive (just as in \cite{maldacena}) to start from the most
 general scale and Lorentz
 invariant effective action made out of $U$, $\del_\al U$ and $\rho_p$ and
 see how much restriction is imposed by the invariance under the
 generalized SCT. Such an action must be of the form
\begin{eqnarray}
S &=& -\int d^{p+1}x U^{p+1} f(z,\rho_p) \comma \\
z &\equiv & {\del_\al U \del^\al U \over U^4} \comma
\end{eqnarray}
where $f(z,\rho_p)$ is an arbitrary function. Applying SCT for $U$, $z$ and
 the measure $d^{p+1}x$, the condition for invariance under SCT is
 worked out as
\begin{eqnarray}
0 &=&\delta S = -2 \int d^{p+1}x U^{p+1}\ep\cdot\del U
{\rho_p\over U^3}
 \left( f -2(z+\rho_p^{-1}) \del_z f\right) \period
\end{eqnarray}
Noting that a shift of $f(z,\rho_p)$ by an arbitrary function of $\rho_p$
 does not spoil the invariance, we get a differential equation for
 $f$, with an arbitrary function $c(\rho_p)$:
\begin{eqnarray}
 f+c(\rho_p)  &=& 2(z+ \rho_p^{-1}) {\del f \over \del z}\period
\end{eqnarray}
Its general solution  is
\begin{eqnarray}
f &=& a(\rho_p)\left(  \sqrt{1+\rho_p z} - b(\rho_p)\right)\comma
\end{eqnarray}
where $a(\rho_p)$ and $b(\rho_p)$ are arbitrary. This is as much as
 GCS dictates on the form of $S$.
\par
The remaining two functions $a(\rho_p)$ and $b(\rho_p)$ can then be
 fixed by invoking the following non-renormalization theorems.
First the BPS condition that there is no
  static force between the D-branes
 fixes $b(\rho_p)$ to be unity.
  Further if the ${\cal O}(z)$ term is not renormalized from the
 simple tree level form, then
  $a(\rho_p)$ is determined to be equal to $ (Nd_p/(2\pi)^2)\rho_p^{-2}$.
Altogether, we get the familiar DBI action
\begin{eqnarray}
S_{DBI} &=& -\int d^{p+1}x {1\over (2\pi)^2 g^2}{U^{7-p}\over Q_p}
 \left( \sqrt{1+ Q_p{\del_\al U \del^\al U \over U^{7-p}}}-1\right)\period
\end{eqnarray}
Thus, it should now be clear that GCS for general $p$ is
 just as powerful as the usual conformal symmetry for $p=3$.
\subsection{GCS for classical super Yang-Mills}
We now turn to the $(p+1)$-dimensional super Yang-Mills theory
describing the low energy dynamics of near-coincident $N$
 D$p$-branes\footnote{Although
 the discussions to follow will go through formally for any
 $p$, we shall restrict ourselves to $0\le p \le 3$, since the quantum
 property of the theory above  4-dimensions is not well-understood.}.
Such a theory can be obtained  most simply by the dimensional reduction
 of  ${\cal N}=1$ $U(N)$ 10-dimensional super Yang-Mills theory, the
 classical action of which is give by
\begin{eqnarray}
S_{10} &=& \int d^{10}x\, \Tr \left\{ -{1\over 4g_{10}^2}
 F_{MN}F^{MN} + {i\over 2} \bar{\psi} \Ga^M
 \left[D_M,\psi\right] \right\} \comma \\
D_M &=& \del_M -iA_M \period
\end{eqnarray}
It is not difficult to check that the fermionic part of the action,
including its reduction, is invariant under the usual conformal
transformations in any dimensions provided appropriate
dimensions for
the fermion fields are assigned.
Therefore, we will concentrate on
 the bosonic part. When reduced to $(p+1)$-dimensions, it takes the
 form
\begin{eqnarray}
 S_{bosonic} &=&  \Tr \int d^{p+1} x \left\{
 -{1\over 4g^2} F_{\mu\nu}F^{\mu\nu}
 -{1\over 2g^2} D_\mu X_m D^\mu X^m
 + {1\over 4g^2} \com{X_m}{X_n} ^2  \right\}\comma
\end{eqnarray}
where  the Greek (Latin) indices run in the range $0\sim p$ ($p+1 \sim 9$)
 and $X_m$ are the Higgs scalars.
\par
Let us  describe the generalized conformal symmetry possessed by
 this action. Consider first the usual conformal transformations of
 the relevant fields, especially the dilatation $\delta^D_\ep$
 and SCT $\de$. The
 variations {\it at numerically the same point $x$},
 which are more convenient
 in the following, are
\begin{eqnarray}
&& \qquad \delta^D_\ep X_m = -(\ep +x\cdot \del)X_m \qquad
 \delta^D_\ep A^\mu = -(\ep +x\cdot \del)A^\mu \comma \\
&& \qquad \de X_m = -2\ep\cdot x X_m(x) -(\de x^\al)\del_\al X_m
 \comma \\
&& \qquad \de A^\mu = -2 \ep \cdot x A^\mu -2(x\cdot A \ep^\mu
 -\ep \cdot A x^\mu ) -(\de x^\al)\del_\al A^\mu \comma
\end{eqnarray}
where the SCT variation $\de x^\al$ for the coordinate is of
 the \lq\lq canonical" form
\begin{eqnarray}
\de x^\al &=& 2\ep\cdot x x^\al -\ep^\al x^2 \period \label{candex}
\end{eqnarray}
Because of the presence of the coupling $g^2$,
the action $S_{bosonic}$ is not
  invariant under these transformations, except for $p=3$.
However, just as in the case of the supergravity description of the
 D$p$-brane system discussed in the previous subsection, we can make it
 invariant if we regard the coupling $g^2$ as a background field $g^2(x)$
 transforming like a scalar field of mass-dimension $3-p$, namely,
\begin{eqnarray}
\delta^D_\ep g^2 &=& -\ep( 3-p +x\cdot \del) g^2 \comma \\\
\de g^2 &=& -2(3-p)\ep\cdot x g^2 -(\de x^\al)\del_\al g^2 \period
\label{degsq}
\end{eqnarray}
The proof is a straightforward exercise.
\par
Thus, we have shown that indeed the concept of GCS can be extended
 to the system of D$p$-branes for general $p$ both on the supergravity side
 and on the super Yang-Mills side.
The notable difference in the form of SCT on two sides, however, exists
 just like  in the case of
  ordinary conformal symmetry for D3-brane system.
 In the next section, we shall clarify the nature of this phenomenon
 and give the precise correspondence.
\section{Relation between the Realizations of GCS  in
   Super Yang-Mills  and  in  Supergravity}
\subsection{Quantum form of GCS for super Yang-Mills}
As was briefly reviewed in the Introduction, the apparent gap between
 the SCT laws in the supergravity and the Yang-Mills theory can be shown
 to be neatly filled by a quantum effect on the Yang-Mills side
  in the case of  D3-brane system \cite{jky}.
It is then natural to expect that the same mechanism should be at work
 in the case of GCS as well. It turns out, however, that there is a subtle
 but important difference between the two categories.
\par
Let us first apply the logic of \cite{jky} to the D$p$-brane system
 for general $p$ and see what happens.
Actually, instead of generalizing the argument of \cite{jky}
 directly,  we will use
 a more systematic BRST approach developed by Fradkin and Palchik
 \cite{fradkin-palchik}, which is suitable for dealing with the
standard background gauge.
Let us decompose the Higgs field $X_m$  as
\begin{eqnarray}
X_m &=& B_m + Y_m \comma
\end{eqnarray}
with  $B_m$  the diagonal background and  $Y_m$  the quantum fluctuation.
We take the gauge-fixing and the corresponding ghost actions to be
(hereafter $D=p+1$)
\begin{eqnarray}
S_{gf} &=& -\half \int {d^D x \over g^2} \Tr\, G^2 \comma \\
G &=& -\del_\mu A^\mu + i\com{B_m}{Y_m}  \comma \\
S_{gh} &=& i\int d^Dx \Tr\, \left( -\cbar\del^\mu D_\mu C
 + \cbar\com{B_m}{\com{X_m}{C} }  \right)\period
\end{eqnarray}
The total action is invariant under the BRST transformation,
 with a fermionic parameter $\lam$,
\begin{eqnarray}
\delta_B X_m &=& -i\com{C}{X_m} \lam \comma \qquad
 \delta_B A_\mu = -\com{D_\mu}{C} \lam \comma \\
\delta_B C &=& iC^2 \lam \comma \qquad \delta_B \cbar = {i\over g^2}
 G\lam \period
\end{eqnarray}
Now let us apply the generalized conformal transformations. $C$ and $\cbar$
 will be regarded as scalar fields with dimension $0$ and $D-2$
 respectively.
Then one finds that, while the scale invariance is trivial, $S_{gf}$ and
 $S_{gh}$ are not invariant under SCT:
\begin{eqnarray}
\delta_\ep S_{gf} &=& 2(D-2) \int {d^{D}x \over g^2}
 \Tr G A\cdot \ep \comma \label{sctgf}\\
\delta_\ep S_{gh} &=& -2(D-2)i \int d^{D}x \Tr \cbar \ep^\mu D_\mu C
\period \label{sctgh}
\end{eqnarray}
A remarkable fact is that one can find a compensating field-dependent
 BRST transformation which removes these unwanted variations.
Take the fermionic parameter $\lam$ to be
\begin{eqnarray}
\lam &=& -2(D-2) \int d^{D}y \Tr ( \cbar(y) A(y)\cdot \ep ) \comma
\end{eqnarray}
and denote this special BRST variation by $\Delta_\ep$.
Then the total action is invariant but the functional measure
 undergoes a non-trivial transformation. It is easy to find
\begin{eqnarray}
{\cal D} (A+\Delta_\ep A) {\cal D}(\cbar + \Delta_\ep \cbar)
 &=& {\cal D}A {\cal D}\cbar \exp \left( i ( -\de S_{gf}
 -\de S_{gh})\right) \comma
\end{eqnarray}
where $\de S_{gf}$ and $\de S_{gh}$ are precisely as given in
(\ref{sctgf}) and (\ref{sctgh}). (Measures for other fields are invariant.)
\par
We have now established the precise form of {\it quantum} GCS :
\begin{itemize}
\item {\it
 Super Yang-Mills theory for  D$p$-brane system is invariant
 under the generalized conformal symmetry, with the modified SCT
 given by }
\begin{eqnarray}
\dte &=& \de + \Delta_\ep \period \label{qgcs}
\end{eqnarray}
\end{itemize}
In particular this leads to the Ward identity for the effective action
$\Ga[B,g^2]$, which is 1PI with respect to the background field $B$:
\begin{eqnarray}
\int d^{p+1}x \left( \de g^2(x) {\delta \over \delta g^2(x)}
 +(\de B(x)  + \Delta_\ep B(x)) {\delta \over \delta B(x)}
\right) \Ga[B,g^2] &=& 0 \period
\end{eqnarray}
We emphasize that this is an exact statement for any background.
\subsection{ GCS at leading order}
Following the formalism developed above,
let us compute the extra piece $\Delta_\ep B$ of the SCT
explicitly. Denote by $B_{m,i}$
 the i-th diagonal component of $B$.  The extra contribution
 is given by
\begin{eqnarray}
\Delta_\ep B_{m,i} &=& 2i(D-2) \langle \com{C}{X_m} _{ii}
 \int d^{D}y \Tr (\cbar(y) A(y)\cdot \ep ) \rangle \comma
\label{modsct}
\end{eqnarray}
where $\langle \quad \rangle$ denotes the expectation value.
For ease of calculation, go to the Euclidean formulation by making
 the following replacements:
$ d^{D}y = -i d^D \ytil$, $ A(y)\cdot \ep = \Atil \cdot \eptil$,
$ \Atil_0 = -i A_0$, $ \eptil^0 = i\ep^0 $. We will be interested
 in the correction which is leading order in the velocity $\del B$.
Then  (\ref{modsct}) can be approximated by
\begin{eqnarray}
\Delta_\ep B_{m,i}(\xtil) &=& 2(D-2)
 \int d^{D}\ytil \Biggl\{
  \langle C_{ij}(\xtil)\cbar_{ji}(\ytil)\rangle \langle Y_{m,ji}(\xtil)
 \Atil_{\mu,ij}(\ytil)\rangle \eptil^\mu
 -(i\leftrightarrow j) \Biggr\}\period
\end{eqnarray}
To the same order of accuracy, the relevant 2-point functions are given
 by
\begin{eqnarray}
\langle C_{ij} (\xtil) \cbar_{ij}(\ytil)\rangle &=&
 i \langle \xtil | \Delta_{ij} | \ytil \rangle \comma \\
\langle Y_{m,ji}(\xtil) \Atil_{\mu,ij}(\ytil)\rangle
 &=& -2i \del_\mu B_{m,ij} g^2  \langle \xtil | \Delta_{ij}^2 |
 \ytil \rangle \comma \\
B_{m,ij} &\equiv & B_{m,i} -B_{m,j} \comma
\end{eqnarray}
where the basic propagator is $\langle \xtil | \Delta_{ij} | \ytil \rangle
 = \int d^D p / (2\pi )^D ( p^2 + B_{ij}^2)^{-1}
 e^{ip\cdot (\xtil -\ytil)}$ and $B_{ij}^2$ is defined as
 $B_{ij}^2 = \sum_m B_{m,ij}^2$.
 Using the formula
\begin{eqnarray}
\calI^{D}_n(B_{ij}) &\equiv &
\langle \xtil | \Delta_{ij}^n | \xtil \rangle =
\int {d^D p \over  (2\pi )^D} {1\over (p^2 + B_{ij}^2)^n}
= { \Ga(n-(D/2)) \over (4\pi)^{D/2} \Ga(n)\,  B_{ij}^{2n-D}} \comma
\label{propn}
\end{eqnarray}
and going back to the Minkowski notation, we get
\begin{eqnarray}
\Delta_\ep B_{m,i} &=& \sum_j {4(D-2) \Ga(3-(D/2))g^2 \over (4\pi)^{D/2}
\, B_{ij}^{6-D}} \ep \cdot \del  B_{m,ij} \period \label{deb}
\end{eqnarray}
Let us specialize to the typical source-probe situation with $N$ D$p$-branes
 as the source at the origin and $B_m$ the probe coordinate.
Taking into account the
 relation between  $B\equiv \sqrt{\sum_m B_m^2}$ and  the supergravity
 coordinate $U$, namely,  $U = 2\pi B$, the formula (\ref{deb}) for this
 configuration becomes
\begin{eqnarray}
\Delta_\ep U &=& {p-1 \over 2} {k\rho_p \over U^2} \ep \cdot \del U \period
\end{eqnarray}
Remember that we have been using the scheme in which  the variation is
 taken at the same point with the underlying canonical
 transformation (\ref{candex}). Therefore, if we convert to the scheme where
 $U$ is transformed canonically without $\Delta_\ep U$ piece
 as in the supergravity treatment, the
 coordinate transformation should be taken as
\begin{eqnarray}
\de x^\al &=& 2\ep\cdot xx^\al -\ep^\al x^2 - {p-1\over 2} \ep^\al
 {k\rho_p \over U^2} \period \label{dexym}
\end{eqnarray}
This agrees  with the AdS-type transformation law (\ref{dexads}) only
 when  the factor $(p-1)/2$ equals unity, \ie  for $p=3$!
This conforms to our previous result \cite{jky} for
 $p=3$, but it is quite puzzling. On one hand,
 the GCS as formulated in
(\ref{qgcs}) must certainly be the symmetry of the effective action
 for super Yang-Mills for any $p$ and hence (\ref{dexym}) should
 be the correct transformation law. On the other hand, at least
for the D0-brane system, the Yang-Mills effective action has been
 checked to agree with  the DBI action  to 2-loop order  \cite{becker-becker}
 and the latter is invariant under (\ref{dexads}), not under
(\ref{dexym}) with $p=0$.  We shall resolve this
apparent contradiction in the next two subsections.
\subsection{Examination of 1-loop effective action}
      From the point of view of Yang-Mills theory, 
the key to the resolution
 of the puzzle lies in the careful treatment of the
 coordinate dependence of the coupling $g(x)$.
In computing the effective action itself, we must carefully keep
 terms linear in $\del g$, which are neglected in the usual calculation.
Under SCT defined in (\ref{degsq}), $\del_\al g$ transforms
 like $\de \del_\al g = -(3-p)\ep_\al g + {\cal O}(\del g) $ and
 produces a finite contribution even as we set $\del g$ to zero after
 the transformation.
\par
Let us then investigate how the effective action is modified due
 to this effect.
For simplicity of presentation, we exhibit the D0-brane case in some
 detail. Extension to general $p$ is entirely straightforward.
In the Euclidean formulation, the total action $\Stil$
 for the D0-brane system takes the form\footnote{When dealing with
 the D0-brane system, for simplicity
we shall use the often-adopted scheme; namely, we
 rescale $X$ by a factor $2\pi \al'$ so that
 it carries  the dimension of length and then set $l_s =\sqrt{\al'}
 =1$.}
\begin{eqnarray}
\Stil &=& \int d\tau \Tr \Biggl\{ {1\over 2g_s}(D_\tau X_m)^2
 - {1\over 4g_s}\com{X_m}{X_n} ^2  \nn\\
&& -\half \theta^TD_\tau\theta -\half \theta^T \ga^m\com{X_m}{\theta}
\Biggr\} + \Stil_{gf} + \Stil_{gh} \comma \\
\Stil_{gf} &=& \int {d\tau \over 2g_s}\Tr
\left(  -\del_\tau \Atil + i\com{B_m}{X_m}\right)^2 \comma  \\
\Stil_{gh} &=& i\int d\tau \Tr \Bigl\{
 \cbar \del_\tau D_\tau C - \cbar \com{B_m}{\com{X_m}{C}} \Bigr\} \period
\end{eqnarray}
We will be interested in the dependence linear in the quantity
\begin{eqnarray}
\eta(\tau ) &\equiv & {\del_\tau g_s \over g_s} \period
\end{eqnarray}
Since the details of the 1-loop calculation  with
 constant $g_s$ is well-documented (see for example  \cite{becker-becker})
 we shall only indicate the modification due to the presence
 of $\eta(\tau)$. When expanded about the background field,
the  quadratic parts which are modified at $\calO(\eta)$ are
\begin{eqnarray}
\calL_{YY} &=& {1\over 2g_s} Y_{m,ij}(-\del_\tau^2
 + \eta(\tau)\del_\tau + B_{ij}^2)Y_{m,ji}\comma  \\
\calL_{\Atil\Atil} &=& {1\over 2g_s} \Atil_{ij}(-\del_\tau^2
 + \eta(\tau)\del_\tau + B_{ij}^2)\Atil_{ji}\comma  \\
\calL_{Y\Atil} &=& {2i\over g_s} \left( \dot{B}_{m,ij}
 -\half \eta(\tau) B_{m,ij}\right) Y_{m,ij} \Atil_{ji}\comma
\end{eqnarray}
where $B_{m,ij}$ and $B_{ij}$ are as defined previously.
Fermions and ghosts are not affected.
$Y\Atil$-mixing can be analyzed in exactly the same way as for the
 constant $g_s$ case if we make the following replacement:
\begin{eqnarray}
 \dot{B}_{ij} & \rightarrow & V_{ij} \comma \\
V_{ij} &=& \left[ \sum_m  \left( \dot{B}_{m,ij}
 -\half \eta(\tau) B_{m,ij}\right)^2\right]^{1/2} \nn\\
& = & \dot{B}_{ij} -\half \eta {\sum_m \dot{B}_{m,ij}B_{m,ij}
 \over \dot{B}_{ij}} + \calO(\eta^2) \period
\end{eqnarray}
Then the Euclidean 1-loop effective action  can  be computed as
\begin{eqnarray}
e^{-(\Ga_1+\Delta \Ga_1)}
&=& \prod_{i<j} \det \left( -\del_\tau^2 + \eta\del_\tau
 + B_{ij}^2\right)^{-8} \nn\\
&& \times \det \left( -\del_\tau^2 + \eta\del_\tau  + B_{ij}^2
 + 2V_{ij}\right)^{-1} \nn\\
&& \times \det \left( -\del_\tau^2 + \eta\del_\tau  + B_{ij}^2
 -2V_{ij}\right)^{-1} \nn\\
&& \times \det \left( -\del_\tau^2 +B_{ij}^2 +\dot{B}_{ij} \right)^4 \nn\\
&& \times  \det \left( -\del_\tau^2 +B_{ij}^2 -\dot{B}_{ij} \right)^4 \nn\\
&& \times \det \left( -\del_\tau^2 +B_{ij}^2\right)^2 \comma
\end{eqnarray}
where $\Ga_1$ is the usual contribution and $\Delta \Ga_1$ is the extra
 part linear in $\eta(\tau)$.
For simplicity, we will drop the subscripts $(ij)$ and make the
 eikonal approximation. Namely  we set
\begin{eqnarray}
x &\equiv & B = b + \vtil \tau = b + vt \comma \label{defx}\\
b\cdot \vtil &=& 0 \comma \\
r^2 &\equiv & B\cdot B = b^2 + \vtil^2 \tau^2 = b^2 + v^2 t^2\comma
\label{defr}
\end{eqnarray}
where $\vtil(v)$ is the Euclidean (Minkowski)
velocity\footnote{ The  condition
in the second line   can always be achieved by a constant shift of $\tau$
 and a redefinition of $b$.}. Then, keeping terms linear in $\eta(\tau)$
 and expanding up to ${\cal O}(\vtil^4)$, $\Delta\Ga_1$ is given by
\begin{eqnarray}
\Delta \Ga_1 &=& 10\Tr\left\{ \Delta \eta\del_\tau
-\vtil^2\Delta \eta\del_\tau \Delta \tau^2
 + \vtil^4 \Delta \eta \del_\tau \Delta\tau^2 \Delta \tau^2\right\}\nn\\
&& + 4 \Tr\Biggl\{ \vtil^2
 ( \Delta ^2 \eta \tau + 2\Delta^3 \eta\del_\tau )
 + \vtil^4 ( -2\Delta \tau^2 \Delta ^2 \eta \tau
 -6 \Delta^3 \eta \del_\tau \Delta \tau^2
 + 4\Delta^4 \eta\tau ) \Biggr\} \comma \nn\\
\end{eqnarray}
where $\Delta \equiv (-\del_\tau^2 + b^2)^{-1}$ is the basic
 propagator.  The calculation of the trace is a bit tedious but
 straightforward using the integration formula (\ref{propn}) with
 $D=1$. The leading contribution, after converting back to Minkowski
 space and supplying a factor of $N$ for the source-probe situation, turned
 out to be
\begin{eqnarray}
 \Delta \Ga_1 &=& N\int dt \eta(t) t
 \left( -6v^2 \calI^{1}_2(b) - 12 v^4 \calI_4^1(b)\right) \label{dg1int}\\
&=& N\int dt {\dot{g}_s \over g_s}  \left( -{3\over 2}{v^2t \over b^4}
 -{15\over 8}{v^2 v^2t\over b^7}\right) \period
\end{eqnarray}
Assuming that the subleading corrections can be taken into account
 by the replacements $b \rightarrow r$, $v^2t \rightarrow v\cdot x$
(see (\ref{defx}) $\sim$ (\ref{defr})), this  corresponds to
\begin{eqnarray}
 \Delta \Ga_1 &=& N \int dt {\dot{g}_s \over g_s}
 \left( -{3\over 2}{v\cdot x \over r^3}
 -{15 \over 8}{v^2 v\cdot x \over r^7} \right) \period
\end{eqnarray}
\par
Having obtained the correction, let us see how it varies under SCT.
In what follows, we will omit the infinitesimal parameter $\ep_0$ for
 simplicity, and denote the SCT variation by $\dk$. Then, the SCT's
at fixed $t$  for the D0-brane case can be written as
\begin{eqnarray}
\dk x &=& 2tx + t^2 v \comma \qquad
 \dk r = 2tr + t^2 {v\cdot x \over r} \comma \qquad
\dk v = 2x + 4tv + t^2\dot{v}  \comma \label{sctd0}\\
 \dk {\dot{g}_s\over g_s} &=& 6 + 2t{\dot{g}_s \over g_s} + t^2
 {d\over dt}\left( {\dot{g}_s \over g_s}\right) \period
\end{eqnarray}
As was already explained before, we may set $\dot{g}_s$ to zero after
 SCT is made. Thus  for $\Delta \Ga_1$ we only need to use the last of
these formulae in the form  $\dk (\dot{g}_s/g_s) =  6$. Then since
$-v\cdot x/r^3 = (d/dt) r^{-1}$, the variation of the first term in $\Delta
 \Ga_1$ becomes a total derivative and can be dropped, and we get
\begin{eqnarray}
 \dk\Delta\Ga_1 &=& \int dt {-45N \over 4} {v^2 v\cdot x \over r^7}
\period \label{sctdg1}
\end{eqnarray}
On the other hand, the usual 1-loop effective action $\Ga_1$ is
\begin{eqnarray}
\Ga_1 &=& {15 N \over 16} {v^4 \over r^7} \comma
\end{eqnarray}
and its SCT variation, computed using (\ref{sctd0}), takes the form
\begin{eqnarray}
\dk \Ga_1 &=& \int dt {30 N \over 4} {v^2 v\cdot x \over r^7}\period
\label{sctg1}
\end{eqnarray}
\noindent    From (\ref{sctdg1}) and (\ref{sctg1}), we find
\begin{eqnarray}
\dk (\Ga_1 + \Delta\Ga_1) &=& -\half \dk \Ga_1
= \left({p-1 \over 2}\right)_{p=0}\dk \Ga_1 \period
\label{loopvariation}
\end{eqnarray}
This shows  that indeed, with the proper correction
proportional to $ \dot{g}_s$,
the appearance of the extra factor $(p-1)/2$ for the SCT
 variation predicted by GCS is realized in D0-brane Yang-Mills theory.
It is easy to confirm that the variation (\ref{loopvariation}) is precisely
cancelled by the variation of  the lowest order action $\Gamma_0\equiv \int dt {v^2\over 2g_s}$ arising from the  $U$-dependent modified term in the
loop-corrected SCT (\ref{dexym}).  
\par
Demonstration of such a consistency for the D$p$-brane super Yang-Mills theory
 for general $p$ is entirely similar:
 Essentially, the only difference from the D0-brane case is the use of
 $p$-dimensional integrals ${\cal I}^p_2$ and ${\cal I}^p_4$,
 defined in (\ref{propn}),  in place
 of ${\cal I}^1_2$ and ${\cal I}^1_4$ in the formula corresponding to
 (\ref{dg1int}). In this way one obtains
\begin{eqnarray}
\Ga_1 + \Delta\Ga_1 &=& N\int d^{p+1}x \left( {C\over 8}
 {(\del B \cdot \del B)^2 \over B^{7-p}}
 -{C \over 4} {(\del B)^2 \del_\al B \over B^{6-p}}
 {\del_\al Q_p \over Q_p} \right) \comma \label{g1fordp}\\
C &=& 2^{2-p} \pi ^{-(p+2)/2} \Ga\left( {7-p \over 2}\right) \comma
\end{eqnarray}
where the second term in (\ref{g1fordp}) represents the correction
 $\Delta \Ga_1$.
By using the SCT previously defined for general $p$, one can easily
 check that the relation $\de (\Ga_1 + \Delta \Ga_1) =
 {p-1\over 2} \de \Ga_1$ is  satisfied, and hence  also that the total effective action
including the lowest order term is invariant under the modified SCT (\ref{dexym}).
\subsection{Correspondence between super Yang-Mills and supergravity}
So far we have succeeded in solving just about half of the puzzle: When
 carefully analyzed,  the  seemingly mysterious extra factor $(p-1)/2$
in the SCT on the Yang-Mills side is entirely consistent with GCS.
In what follows, we shall solve the other half of the puzzle, namely
 how such a modified effective action is related to the DBI action
 produced in supergravity, in two complementary ways.
Our analysis  will
 disclose an important aspect of the correspondence between supergravity
and Yang-Mills matrix models for general $p$ in
 a rather explicit manner.
\par
First we approach from the super Yang-Mills side. A crucial observation
 that relates the different-looking effective actions will be  that the
 correction $\Delta\Ga_1$ found by the 1-loop calculation can be
 reproduced by a simple redefinition of the probe coordinate in the
 usual form of the effective action. Actually, this idea  naturally
 emerges in the  effort to understand the correction term in the
 effective action from  the supergravity side. So let us present this
 reasoning before we write down the precise field redefintion to be made.
\par
For simplicity, consider the D0-brane system.
Recall that, from the 11 dimensional viewpoint,
the effective action for a probe D-particle in the
field of a cluster of fixed source D-particles is given by  \cite{bbpt}
\begin{equation}
S_0=
-\int d\tau \, p_-{dx^-\over d\tau} .
\end{equation}
where ${dx^-/ d\tau}$ is determined from the massless constraint
\begin{equation}
g_{\mu\nu}(x(s) )\,
{dx^{\mu}(s)\over ds}{dx^{\nu}(s) \over ds} =0 .
\label{masslesscondition}
\end{equation}
Our convention for the light-cone coordinate is
$x^{\pm}=x^{11}\pm t$,\,
$A\cdot B = {1\over 2}(A^+B^- + A^-B^+) + A_iB_i$,\,
$2A_-=A^+$,\, $ 2A_+=A^-$.
Thus in the linearized approximation for the gravitational field, the general
form of the action takes the form
\begin{equation}
S_D=
\int d\tau \, {p_- \over 2}\Bigl(
\bigl({dx^i\over d\tau}\bigr)^2 +
h_{\mu\nu}(x)s_2^{\mu}s_2^{\nu}
\Bigr) ,
\end{equation}
where $s_2^{\mu}$ is the velocity vector of the probe in the lowest order
approximation,  defined as
$(s_2^+, s_2^-, s_2^i)=(2, -{1\over 2}v^2, v_i)$.
For constant $g_s$, the interaction term
comes from the component $h_{--}=N_1\kappa_{11}^2{15\over 2 r^7}$, which
 is the solution of the linearized Einstein equation
$-{1\over 2}\triangle h_{\mu\nu}
= \kappa_{11}^2 T_{\mu\nu}$ with the
 the energy-momentum tensor
\begin{equation}
T_{\mu\nu}(x)= {N_1\over 2\pi R^2}\delta^9(x_{\perp})s_{1\mu}s_{1\nu}
\end{equation}
for the source at rest, namely for
 $(s_1^+, s_1^-, s_1^i)=(2s_{1-}, 2s_{1, +}, s_1^i) =
(2, 0, 0)$.
Now for $(s_2^+, s_2^-, s_2^i)$ given above, the general form
 of $h_{\mu\nu}(x)s_2^{\mu}s_2^{\nu}$ is
\begin{eqnarray}
h_{\mu\nu}(x)s_2^{\mu}s_2^{\nu}
=4h_{++} - 2h_{+-}v^2  +{1\over 4}h_{--}v^4 +4h_{+i}v^i
-h_{-i}v^2v^i +
h_{ij}v^iv^j\period
\end{eqnarray}
Recall that the relevant correction term containing $\dot{g}_s$
 is of order $v^3$.  Combining with  the fact
 that the gravitational field produced
by the fixed source does not depend on the velocity of the
probe, we see that the only possibility for producing such a
 term in the effective action would be  to have
 $h_{-i}\propto \dot{g}_s x^i /r^7$. This, however, is impossible
since, in the linearized approximation,
  the energy-momentum tensor of the source
cannot have the $-i$ component. After all,
it is difficult to imagine that the time-dependence
of the dilaton would produce such a
component for the energy-momentum  tensor for the
point-like source D-particle at rest.
\par
Blessedly, there is a way out. Suppose we make a
redefinition of the time variable $t \rightarrow t+ \phi(x) $
for the probe D-particle, where $\phi(x)$ is a function
of the spatial coordinate of the probe. Then, this induces a
 shift of the metric
\begin{eqnarray}
h_{-i} \rightarrow h_{-i} + \partial_{i} \phi\comma
\end{eqnarray}
which provides the missing component $h_{-i}$ capable of
 accounting for the  modified term,  with
$\phi(x)\propto \dot{g}_s / r^5$.
This strongly suggests  that, when $g_s$ is not constant,
the collective coordinate for the probe D-particle
must be appropriately chosen in order for the
motion of the D-particle to  be described by the
standard language of supergravity.
\par
This leads us, then, to consider the following replacement
 of the probe coordinate
\begin{eqnarray}
x \rightarrow x' = x + K\dot{g}_s{v\over r^5} \comma
\end{eqnarray}
with some constant $K$. It induces a change in velocity of the form
\begin{eqnarray}
v \rightarrow v' =v +K\dot{g}_s
\bigl(
{\dot{v}\over r^5} -5{v\cdot x\over r^7}v \bigr)\period
\end{eqnarray}
Then, the tree-level action is transformed into
\begin{eqnarray}
\int dt {v^2\over 2g_s} \rightarrow \int dt {v'^2\over 2g_s}
 = \int dt {v^2\over 2g_s} -{5K \over 2}
\int dt {\dot{g}_s\over g_s}{v^2 v\cdot x\over r^7} \comma
\end{eqnarray}
to the order of interest. We see immediately that the second
 term reproduces the 1-loop modification if we set $K=3N/4$.
Furthermore, while the original Yang-Mills coordinate $x$
 transforms under SCT like
\begin{eqnarray}
 \dk x = 2tx + t^2 v +\left({p-1\over 2}\right)_{p=0} {3g_s Nv \over r^5}
\comma
\end{eqnarray}
the new coordinate $x'$ is easily seen to transform,
to the accuracy of the present one-loop approximation,
just like
 the one in the usual DBI action, without the factor $(p-1)/2$:
\begin{eqnarray}
\dk x' =  2t x' +
t^2 {dx'\over dt}
-{3g_s N\over 2r'^5}v'
+{3N\over 4}6g_s {v'\over r'^5}= 2t x'+
t^2 {dx'\over dt}
+{3g_s N\over r'^5}v' \period
\end{eqnarray}
\par
The same mechanism works for the general case of D$p$-brane system.
The redefinition of the diagonal Higgs field is
\begin{eqnarray}
X_m &\rightarrow & X'_m = X_m -
{1\over 4}{k \del_\al Q_p \over U^{5-p}}\del_\al X_m \comma
\end{eqnarray}
and again $X'_m$ can be shown to transform like the transverse
 coordinate in the DBI action.

\par
Next, let us show that the same conclusion
 can actually be reached from the supergravity side without
 the knowledge of the 1-loop calculation in the super Yang-Mills theory.
Recall that the SCT for the world-volume coordinate $x^\al$ in the
 supergravity solution takes the  non-canonical form
$\de x^\al = 2\ep\cdot x x^\al -\ep^\al x^2 -\ep^\al(kQ_p /U^{5-p})$.
The fact that the corresponding coordinate in the Yang-Mills theory,
 in contrast,  transforms canonically suggests that we should
 identify the latter as a new coordinate $y^\al(x)$ on supergravity side
 which transforms canonically.
To find such $y^\al(x)$, let us set $y^\al = x^\al + \zeta^\al$ and
write down the condition on $\zeta^\al$. It is given by
\begin{eqnarray}
\de \zeta^\al &=& 2\ep\cdot x \zeta^\al + 2(\ep\cdot \zeta x^\al
 -x\cdot \zeta \ep^\al)  +\ep^\al {kQ_p \over U^{5-p}}-\ep^\al \zeta^2
\period
\end{eqnarray}
Apart from the last term non-linear in $\zeta$, this
 transformation law is recognized to be identical to that of
 $\del^\al \chi$, where  $\chi$ is a scalar field of dimension $=-2$
 given by
\begin{eqnarray}
\chi &=& {1\over 4} {kQ_p \over U^{5-p}} \period
\end{eqnarray}
Therefore we get
\begin{eqnarray}
 \zeta_\al &=& {1\over 4} \del_\al \left( {kQ_p \over U^{5-p}}
\right) = {k\over 4} \left( {\del_\al Q_p \over U^{5-p}}
 + (p-5) {Q_p \del_\al U \over U^{6-p}}\right) \nn\\
&\simeq & {1\over 4} {k \del_\al Q_p \over U^{5-p}}\comma
\end{eqnarray}
where we dropped the second term since $ \del_\al U$ is supposed to be small.
Also since we agree to neglect $(\del_\al Q_p)^2$, the omission of
 $\zeta^2$ part is {\it a posteriori} justified. Thus, to the leading order
 we find
\begin{eqnarray}
y^\al &=& x^\al + {1\over 4} {k \del_\al Q_p \over U^{5-p}} \period
\end{eqnarray}
\par
If we use $y^\al$ as our coordinate, we must then regard the transverse
 coordinate  $X'_m(x)$  of the probe D$p$-brane as a function of $y$. Then
\begin{eqnarray}
X'_m(x) &=& X'_m(y-\zeta) = X'_m(y) -\zeta^\al\del_\al X'_m(y) + \cdots \nn\\
&\simeq & X_m(y)
-{1\over 4}{k \del_\al Q_p \over U^{5-p}}\del_\al X_m(y)\comma \label{relxpx}
\end{eqnarray}
where in the last line we renamed the field by dropping the prime,
 to emphasize that it is considered to be a field different from $X'_m$.
The relation (\ref{relxpx}) is identical in form to the redefinition found
 previously by using  the information obtained from 1-loop calculation,
 except for a slight change in the argument. This difference, however,
 is of higher order in the present approximation. To see this,
let us compute how $X_m(y)$ transforms under SCT {\it at a fixed point}.
 Using $\de X'_m(x) = -2\ep\cdot x X'_m(x) -(\de x^\al + \De x^\al) \del_\al
 X'_m(x) $,
we get
\begin{eqnarray}
\de X_m(y) &=& \de \left( X'_m(x) +
{1\over 4}{k \del_\al Q_p \over U^{5-p}}\del_\al X'_m(x) \right) \nn\\
&=& -2\ep\cdot x X'_m(x) -(\de x^\al + \De x^\al) \del_\al
 X'_m(x) \nn\\
&& \quad + {1\over 4}{k \del_\al \de Q_p \over U^{5-p}}\del_\al X'_m(x)
 + \calO(\del Q_p) \period
\end{eqnarray}
Remembering that after the transformation we may set $\del Q=0$, hence
 $y=x$, we find after a little calculation,
\begin{eqnarray}
\de X_m(x) &=& -2\ep\cdot x X_m(x) -\de x^\al \del_\al \Xtil_m(x)
 + {p-1 \over 2} \ep^\al {1\over 4}
{k \del_\al Q_p \over U^{5-p}}\del_\al X_m(x) \period
\end{eqnarray}
This shows that $X_m$ exhibits precisely the transformation law of
 the Higgs field in super Yang-Mills description.
\par
Thus, we have reached the same result in two complementary
 ways: The realizations of GCS, which strongly controls the D-brane dynamics
 on both  the supergravity and super Yang-Mills side
take apparently different forms for general $p$, but they are
 related by a subtle redefinition of the Higgs field or by a
 corresponding transformation of the world-volume coordinate.
We emphasize that this is not a technical detail. It is of utmost
 importance for the consistency of the powerful concept of GCS.
\par
To summarize, we have shown that, when the non-constancy of the
 coupling is taken into account, the effective action
in general contains terms which depend on the world-volume derivatives
of the coupling-constant fields.
However, there is a  field redefinition
which eliminates the terms of first order in
the derivatives of the coupling-constant fields and make the
SCT laws to be those of the (pseudo) AdS space-times.
Such a redefinition is in fact compulsory, as our first argument
 told  us, in order that the
dynamics of the probe D-brane be described by
supergravity.
It is quite remarkable that
while this effect appears on the Yang-Mills side as a quantum loop effect,
the same field redefinition is dictated  in
the classical structure of supergravity.  This provides us with
 a justification
of  the construction of the effective actions made in section II,
assuming implicitly that there are no terms which are of first order
with respect to the world-volume derivatives of the coupling-constant field.
It also provides yet another fine example supporting the
correspondence between the  classical
supergravity and loop-corrected super Yang-Mills theory for general $p$.

\subsection{Generalized conformal transformation
for general configurations of D-branes}
It should by now be fairly clear  that our method of
deriving the generalized conformal transformation
in super Yang-Mills theory
can be extended  to  backgrounds
of arbitrary configurations of D-branes.
As a matter of fact, study of such an extension turns out to shed
  a further light on the meaning of the generalized conformal symmetry.
 So let us briefly discuss this generalization,
 taking the system of D-particles as the simplest  example.
\par
Consider a system of $n$ clusters, each of which consists
 of a large number of coincident D-particles, and denote by $x_a\
 (a=1,2,\ldots, n)$ and $N_a$  the coodinate and the number
 of D-particles, respectively, of the $a$-th cluster.
The quantum-modified SCT before the field redefinition is
\begin{equation}
\delta_K x_a = 2t x_a+t^2{dx_a \over dt}
-\sum_b{3g_sN_b \over 2r_{ab}^5}v_{ab},
\end{equation}
and the effective action up to the first order in $\dot{g}_s$ is
\begin{equation}
\Gamma=\int dt \Bigl[
\sum_a{N_a\over 2g_s}v_a^2
 + {1\over 2}\sum_{a,b}{15N_aN_ b \over 16}{v_{ab}^4 \over r_{ab}^7}
-{1\over 2}\sum_{a,b}
{15 N_aN_b \over 8}
{\dot{g}_s \over g_s}{v_{ab}^2 v_{ab}\cdot x_{ab} \over r_{ab}^7}\Bigr] .
\end{equation}
The field redefinition $x_a \rightarrow x'_a$ of the D-particle
collective coordinates  which eliminates the dependence on $\dot{g}_s$
 takes the form
\begin{equation}
x_a =x'_a -\sum_b {3N_b\over 4} {\dot{g}_s v_{ab}\over r_{ab}^5} \period
\end{equation}
SCT for the new coordinate $x_a'$ then becomes
\begin{equation}
\delta_K x'_a = 2t x'_a+t^2{dx'_a \over dt}
+\sum_b{3g_sN_b \over r_{ab}'^5}v'_{ab} \period
\label{geneconf}
\end{equation}
It is a simple matter  to check that the usual form of the effective action
with constant string coupling is invariant under the
modified transformation law (\ref{geneconf}) to the accuracy of
 the present approximation.
\par
The transformation law obtained above for the generic
configuration of D-branes exhibits
some notable new features, which were not seen in the special background
often considered, namely that  of a single heavy source consisting of
 coincident D-branes.
\par
First, the transformation law for the generic background no longer
 admits a simple space-time interpretation. 
This is evident from the
 fact that (\ref{geneconf}) involves,  in general, more than
 one relative velocities between D-particles. It means that, at
 least in our present formulation of the multi-cluster system,  an overall
 shift of time cannot  simultaneously be
 responsible  for the redefinitions of coordinates
 for different clusters which move with relative velocities.
This strongly suggests that a proper space-time interpretation of
 the (generalized) conformal transformation for general backgrounds,
 if it is possible at all, would require a completely covariant
 formulation of D-brane dynamics in which a world volume is introduced
 for each D-brane in a reparametrization invariant way.
\par
Secondly, apart from the problem of space-time interpretation, it should
 be pointed out that
 the system with multi-centered heavy D$p$-brane sources does not
 in general  behave in a simple manner under the (generalized)
conformal transformation for any $p$. This is due to the
 fact that (even without the modified term) SCT for the relative
 velocity $v_{ab}$ contains an inhomogeneous term. For instance,
 the relevant transformation law for the D-particle case is
$\delta_K v_{ab} = 2x_{ab}+ 4t v_{ab}+ t^2{d\over dt}v_{ab}
 + \cdots \, $, which involves $2x_{ab}$, not proportional to $v_{ab}$.
 Thus, even if one starts with the sources which are relatively
at rest ($v_{ab}=0$), they inevitably acquire nonzero relative velocities after
the SCT,  except for the case of coincident sources.
 This implies that in order to discuss the conformal symmetry appropriately
in such a general case, we would have to  treat the
 the motions of all the  D-branes on equal footing.
\par
A related question of importance is
how and to what extent the generalized conformal symmetry
 can constrain the many-body dynamics of D-branes.
One possibility, suggested by our discussions in section II, is
that GCS may play an important role in extending
 the non-renormalization theorems, so far checked to 2-loop order for
 special configurations, to higher loops and to general many-body systems.
Since these theorems are believed  to be
 the basis for the correspondence between supergravity and matrix models
for D-branes, at least in the weak coupling region, this question is
 of great interest. For example, it would be important to
examine the general 3-body actions given in \cite{oy} from this
 viewpoint.  This problem will be discussed elsewhere.
\section{Discussions}
In this paper, we have investigated the symmetry structure
 of general  D$p$-brane systems in both their YM and the supergravity
descriptions, with the aim to establish further the
conjectured correspondence between these two seemingly different theories.
As we have already summarized the outcome of this research in the
 Introduction and at appropriate places in the text, we will,
 in the remainder  of this article, offer some further
 observations  which would be of importance in future investigations.
\begin{enumerate}
\item One of the key ideas of our study is to regard the
 coupling constant of the theory, both in the Yang-Mills and in
 supergravity/string descriptions, as a \lq\lq field" which transforms
 non-trivially under the conformal-type transformations. On the Yang-Mills
 side, it  allowed us to extend the notion of conformal symmetry
 that is normally thought to exist  only in 4-dimensions to other
 dimensions. This in turn, through quantum corrections,
 produced in the effective action new terms
 which explicitly depend on the derivative of the coupling constant, and
 they are crucial in reconciling the apparently disparate transformation
 laws between Yang-Mills and supergravity descriptions.
 While we have demonstrated the agreement including the exact coefficients,
 the full understanding and the interpretation of the new terms
 and the notion of dynamical YM coupling constant remains
 to be given. On the supergravity side, (up to reparametrizations)
 the extra degrees of freedom and the induced terms
 in the effective Lagrangian are associated with the
 space-time dependent  dilaton field, which is reflected in the variable
 radius of the AdS-like space-time.
 Thus a more precise description and comparison of
 this phenomenon would require proper inclusion of the dynamics of the dilaton
 degrees of freedom in supergravity.
\item The quantum effect, which was so important in connecting
 the Yang-Mills and the supergravity descriptions, has been
 computed only in the weak coupling regime at one loop. Nevertheless,
 our result robustly supports the conjectured relation between
  supergravity and  the Yang-Mills matrix models.
 Since the Maldacena's original conjecture is supposed to hold at
 large $g^2 N$, this means that there must exist  significant
non-renormalization theorems at work in the weak-coupling region,
 which protect the coefficients of the generalized conformal
 transformation obtained in one-loop calculation as we go to the
 strong-coupling regime. Understanding of these theorems and their
 relation to GCS is an important future problem.
\item In relation to the effects of higher loops, we wish to point out
 the strong similarity of our GCS Ward identities, which explicitly
 involves the derivative with respect to the coupling constant,
  and the standard  renormalization group (RG) equations.  RG equations
 are capable of organizing and summing the infinite series of logarithms
 and as a result the coupling constant is turned into an effective
 running coupling. In particular in asymptotically free theories,
 the essential nature of such a running coupling is determined at
 short distance by the one-loop effect. The fact that our one-loop
 calculation gives the correct coefficients for the modified terms
  in the generalized conformal transformation in the bulk is analogous
 to this phenomenon and suggests that a similar mechanism is operating
 in the present case. This analogy may be of importance in further
 elucidating the structure of the GCS formulated in this work.
\item Among the possible applications of GCS,
 calculations of various correlation functions would be an important
 challenge.
In general,  conformal transformations imply constraints on
correlation functions and in fact they completely specify
 the form of lower point
functions. We expect that the generalized transformations
established in the present work are similarly useful.
 We have already seen the effectiveness of the symmetry
in deriving the form of the
DBI action for $p$-brane probes in a
fixed source. We emphasize that since our
transformations contain in a nonlinear manner the extra radial dimension,
they probe the bulk of the supergravity background.
The generalized conformal
transformations  are then expected to be
very useful for addressing questions such as the bulk
 to boundary or bulk to bulk correlators from the symmetry point of view.
Since at present time almost all the comparisons in the literature
between the bulk and the boundary using the AdS/CFT
correspondence involve only
correlations and sources at the boundary,  the extension to
 the full bulk region is of major
 interest. It would also be useful for
 \lq\lq proving"  the correspondence between the  correlation functions
in the bulk and on the boundary from within the logic intrinsic to
 YM theory.
\item As we have already suggested in subsection IIIE,
 to obtain  a proper
 space-time picture of GCS for general backgrounds, some reparametrization
 invariant formulation on the Yang-Mills side is likely to be
 required. In this regard, we wish to mention a suggestive structure
 already seen in the Yang-Mills action. Take for example the case
 of D-particles. It is not difficult to see that by scaling the
(time-dependent) coupling in the kinetic and the potential terms
appropriately  one can obtain a form that exhibits (time) reparametrization
symmetry.  In this picture the usual Yang-Mills theory with a constant
 coupling may be regarded as a gauge-fixed version. This suggests
 that an extension of the theory with
 symmetry structure encompassing reparametrization of D-brane
 world volume as well as the GCS advocated in this work
 may indeed  be possible.
\end{enumerate}
We hope to address these and some other related problems in  future
 publications.

\vspace{0.4cm}
The work of A.J. is supported in part
by  the Department of Energy under contract DE-FG02-91ER40688-Task A.
The work of Y. K. and T.Y. is supported in part
by Grant-in-Aid for Scientific  Research (No. 09640337)
and Grant-in-Aid for International Scientific Research
(Joint Research, No. 10044061) from the Ministry of  Education, Science and Culture. The final stage of the present work was completed during the visit
of one  (T. Y.) of the authors to Brown University.
T. Y. thanks Phys. Dept. of Brown University
for hospitality during his stay.


\end{document}